\documentstyle[12pt,draft,psfig]{nature-ejb}

\def\la{\ifmmode\stackrel{<}{_{\sim}}\else$\stackrel{<}{_{\sim}}$\fi}
\def\ga{\ifmmode\stackrel{>}{_{\sim}}\else$\stackrel{>}{_{\sim}}$\fi}

\def\ergcms{\mbox{\,erg\,cm$^{-2}$\,s$^{-1}$}}
\def\ergcm{\mbox{\,erg\,cm$^{-2}$}}
\def\sgra{\mbox{SGR~1806--20}}
\def\sgrb{\mbox{SGR~1900+14}}
\def\sgrc{\mbox{SGR~0526--66}}

\def\swift{\textit{Swift}}

\newcommand{\titl}{
Gamma Ray Observations of a \\Giant Flare from the \\Magnetar SGR~1806$-$20
}

\title{\LARGE \bf \titl}
\author{D.~M.~Palmer,$^1$
S.~Barthelmy,$^2$
N. Gehrels,$^2$
R.~M.~Kippen,$^1$
T.~Cayton,$^1$
C.~Kouveliotou,$^3$
D.~Eichler,$^4$
R.~A.~M.~J.~Wijers,$^5$
P.~M.~Woods,$^6$
J.~Granot,$^7$
Y.~E.~Lyubarsky,$^4$
E.~Ramirez-Ruiz,$^8$
L.~Barbier,$^{2}$
M.~Chester,$^9$ 
J.~Cummings,$^{2,10}$
E.~E.~Fenimore,$^1$ 
M.~H.~Finger,$^6$
B.~M.~Gaensler,$^{11}$
D.~Hullinger,$^{2}$
H.~Krimm,$^{2,12}$
C.~B.~Markwardt,$^{2,13}$
J.~A.~Nousek,$^9$
A.~Parsons,$^2$ 
S.~Patel,$^6$
T.~Sakamoto,$^{2,10}$
G.~Sato,$^{14}$
M.~Suzuki$^{15}$ and
J.~Tueller$^2$
\\
\footnotesize
$^1$ Los Alamos National Laboratory, Los Alamos, NM, 87545, USA  \vspace{-2mm} \\
$^2$ NASA/Goddard Space Flight Center, Greenbelt, MD,  20771, USA \vspace{-2mm} \\
$^3$ NASA/Marshall Space Flight Center, NSSTC, XD-12, 320 Sparkman
Dr., Huntsville, AL 35805, USA  \vspace{-2mm} \\
$^4$ Department of Physics, Ben Gurion University, POB 653,
Beer Sheva 84105, Israel \vspace{-2mm} \\
$^5$ Astronomical Institute ``Anton Pannekoek'', University
of Amsterdam, Kruislaan 403, 1098 SJ, Amsterdam, the Netherlands \vspace{-2mm} \\
$^6$ Universities Space Research Association, NSSTC, XD-12,
320 Sparkman Dr., Huntsville, AL 35805, USA  \vspace{-2mm} \\
$^7$ Kavli Institute for Particle Astrophysics and
Cosmology, Stanford University, P.O. Box 20450, MS 29, Stanford, CA
94309, USA  \vspace{-2mm}\\
$^8$ Institute for Advanced Study, Einstein Drive, Princeton, NJ
08540, USA  \vspace{-2mm}  \\
$^9$ Pennsylvania State University, University Park, PA, 16802, USA  \vspace{-2mm} \\
$^{10}$ National Research Council, 500 Fifth St. N.W. Washington, DC, 20001, USA   \vspace{-2mm}\\
$^{11}$ Harvard-Smithsonian Center for Astrophysics,
60 Garden Street MS-6, Cambridge, MA 02138, USA   \vspace{-2mm}  \\
$^{12}$ Universities Space Research Association, Goddard Space Flight Center , Greenbelt, MD,  20771, USA  \vspace{-2mm} \\
$^{13}$ University of Maryland, College Park, MD, 20742, USA  \vspace{-2mm} \\
$^{14}$ Institute of Space and Astronautical Science (ISAS) / JAXA,
 3-1-1 Yoshinodai, Sagamihara, Kanagawa 229-8510, Japan \vspace{-2mm}\\
$^{15}$ Saitama University, 55  Shimo-Okubo, Sakura-ku, Saitama City, Saitama 338-8570, Japan \vspace{-2mm}
}

\dates{--}{--}
\headertitle{Gamma-ray observation of a giant flare}
\mainauthor{Palmer et al.}

\begin{document}

\normalsize

\summary{}

\maketitle


{\bf Magnetars\cite{td95} comprise two classes of rotating neutron stars (Soft Gamma
Repeaters (SGRs) and Anomalous X-ray Pulsars), whose X-ray emission is
powered by an ultrastrong magnetic field, $B \sim 10^{15}$\,G.
Occasionally SGRs enter into active episodes producing many short X-ray bursts;
extremely rarely (about once per 50 years per source), SGRs emit a giant
flare, an event with total energy at least $10^3$ higher than
their typical bursts.\cite{mgi+79,fkl96,hcm+99}
Here we report that, on 2004 December 27, \sgra\ emitted the brightest
extra-solar transient event ever recorded, even surpassing the full moon
brightness for 0.2 seconds.  The total (isotropic)
flare energy is $2\times10^{46}$\,erg, $\sim$100 times higher than the
only two
previous events, making this flare a once in a century event. This
colossal energy release likely occurred during a catastrophic
reconfiguration of the magnetar's magnetic field.
Such an event
would have resembled a short, hard Gamma Ray Burst (GRB) if it had
occurred within 40 Mpc, suggesting that extragalactic SGR flares may
indeed form a subclass of GRBs.}

Only two other giant flares have previously been
recorded, one each from \sgrc\ on 1979 March 5\cite{mgi+79,fkl96} and \sgrb\ on 1998
August 27\cite{hcm+99}.
Intense X-ray burst activity from \sgrb\ preceded the 1998 August 27
flare\cite{hkw+99}; there are very few observations of \sgrc\ preceding the
1979 March 5 event.  In the year leading up to the \sgra\ flare, well-sampled
X-ray monitoring observations of the source with the {\it Rossi X-ray Timing
Explorer (RXTE)} indicated that it was also entering a very active
phase\cite{wkg+04}, emitting more frequent and intense bursts and showing
enhanced persistent X-ray emission that was, indeed, a prelude to the
unprecedented giant flare.

On 2004 December 27 the \swift\ satellite\cite{swift} was among a large number
of spacecraft inundated by radiation from \sgra\cite{bgm+04,pbb+04,rhessi}. 
The  Burst Alert Telescope (BAT)\cite{bat} is a $\gamma$-ray 
(15-350\,keV) coded aperture imager on \swift.
Although \swift\ was turned away from the SGR location,
and so the event illuminated the detector from behind,
the flux that passed through the spacecraft and shielding of
the BAT provided excellent measurements of the event.
The BAT light curve (Figure 1, and with more detail in
the {\it supplementary figures}) demonstrates that magnetar giant flares are remarkably
similar: all three start with an initial very short and spectrally hard main spike, followed by
an extended softer tail highly modulated at the neutron star's spin period. 
The bright, main spike lasts $\sim0.5$\,s
and is followed by a tail with $\sim50$ cycles of high-amplitude
pulsations at the known rotation period of \sgra\ (7.56\,s).
In the Dec.~27 event we also notice a 1-s long, flat-topped precursor burst at
142\,s before the main spike.

Several astounding new properties of a magnetar flare are revealed from the
superb time resolution of the BAT. Figure 1b plots the sharp initial rise of the
main spike in time bins of 100$\mu$s, equivalent to the light crossing time of the
neutron star diameter. Prior to the steep rise of the initial spike, the count
rate was rising for 40~ms at a slower rate (shown in {\it supplementary figure} 2)
and had reached roughly 30,000 cps (above a $\sim 9$\,kcps background)
by $t=0$.   At that point it increased by a factor of more than 100 in less than 1.5 ms 
(0.3\,ms $e$-folding rise time).  
This is followed by at least one dip
and continued brightening (additional dips would not be visible due to instrument saturation)
on its way to the peak.
The flare rise is thus resolved for the first time. 
The flux
during the spike, though heavily attenuated,  saturated the BAT modules,
precluding a reliable flux measurement. We have used, therefore, the
SOPA\cite{sopa} and ESP\cite{esp} instruments located on geosynchronous
satellites (see {\it Supplementary Methods\/} Section) to measure the main peak
flux. The SOPA instruments are small silicon detectors designed with
fast event processing to measure the high particle fluxes found in orbit. 
During the peak of the
burst, each detector had a dead-time greater than 50\%, but this level of
saturation can be accurately corrected for. We fit the SOPA data with an
exponential-cutoff power law (finding a characteristic temperature $kT =
0.48(4)$\,MeV and a power-law photon index $-0.2(1)$) and derive a flux of
$5.0(3)$\ergcms\ over an 0.160\,s integration time for 45\,keV to 10\,MeV
photons; the corresponding fluence is $0.80(5)$\ergcm.
This duration and spectral hardness is in the range of 
characteristics
found for the short, 
hard subclass of classic GRBs.\cite{kmf+93}

Since the count rate was significantly lower during the tail, we
were able to model the off-axis illumination and calibrate the flux
and spectroscopy measurements (see {\it Supplementary
Methods\/}). We find the tail of the burst to have an energy
fluence of $1.0(5)\times10^{-3}$ \ergcm\ at photon energies $>
60$\,keV. The spectral fits are consistent with thermalized
spectra with $kT\sim 15-30$\,keV as seen in previous flares,
implying  a comparable energy fluence below our $60$\,keV
threshold. Accounting for  the $10-60$ keV photons, we project the
total tail fluence to be $\ga 2\times10^{-3}$\ergcm, roughly 0.3\%
that of the main peak. For a distance to \sgra\ of
$15d_{15}$\,kpc\cite{ce04}, we then find an (isotropic equivalent)
energy release of $2\times 10^{46}d_{15}^2$\,erg in the spike, and
$5 \times 10^{43}d_{15}^2$\,erg in the tail. Thus, the
isotropic equivalent energy in the initial spike is about two
orders of magnitude larger than that in the other two giant flares,
while the energy in the tail is comparable.
Indeed, 
a radio afterglow\cite{gkg+05}
was detected from this flare with a 
luminosity 500 times that from the
August 27 flare, suggesting a truly 
large difference in the prompt burst energy.  
The consistency of
the tail energies among the three flares is attributable to the
fact that they are limited by the storage capacity of the magnetic
field\cite{td95,td01} and should be as constant from source to
source as the field energy.
Thus, the tail luminosities, which are
expected to be at roughly the magnetically modified Eddington
limit, should also be similar, as observed. The extent of
magnetic reconnection, on the other hand, governs the prompt
energy release during the main spike; this can vary greatly from
one event to the next, even within the same source.

The pulse profile in the tail of the flare just after the main spike features
one large peak and two smaller adjacent local maxima separated by about 1/4 of
a rotation cycle (Figure 2 and {\it supplementary figure} 1).  
The relative intensities of the peaks change during the tail, but
their phases remain fixed, indicating that the field configuration does not
change substantially during the tail and that the released energy comes from the
trapped fireball.

The polar $B-$field of \sgra\ has been calculated\cite{wkg+02} from its spin-down
rate to be $\sim1.6\times 10^{15}$\,G,
corresponding to a external magnetic field energy of $2\times
10^{47}$\,erg, which indicates that at most $10d_{15}^{-2}/f$ such
giant flares can be produced from the star in its lifetime (here $f$ is
the beaming factor).

We used RXTE to measure the spin frequency 
and spin-down rate of the SGR 30 days after the 
flare.\cite{wkg+05} The frequency is
consistent with an extrapolation
of the pre-flare frequency with pre- and post-flare spin-down rates.
Thus, the December 27
flare could not have caused a rapid, lasting change in the spin frequency
greater than $\sim2 \times 10^{-5}$ Hz; this, despite the much
larger apparent burst energy, limits the  frequency change to be at
most comparable to that seen following the August 27
flare\cite{wkv+99}.
The post-flare spin-down rate, $-3.15(9)\times10^{-12} Hz\ s^{-1}$,
although lower
than it was shortly before the flare, is still in its historical range.

The three time scales in the phenomenon, a) the rise time of $\sim
1\;$ms, b) the duration of the hard spike, $\sim 0.5\;$s, and c)
the duration of the tail, several minutes  - are similar for all
three giant flares. These are attributed to the Alfv\'en
propagation times in a) the magnetosphere and b) the star,  and
c) the cooling time of the trapped pair fireball, respectively
\cite{td95,td01}.

Violent energy dissipation can occur anywhere in the magnetically
dominated region, which includes the outer layers of the neutron
star: if an energy of $10^{46}E_{46}$\,erg is dissipated roughly
uniformly in the reconnection region of volume
$10^{18}V_{18}$\,cm$^3$, then matter above the layer at density
$1\times10^8E_{46}/V_{18}$\,g/cm$^3$ will 
have an energy density larger than its gravitational potential and
become unbound. This is
about $10^{24}$\,g, which can be ejected into the magnetosphere at
fractions of the speed of light. Such a mass ejection  (which need
not be isotropic)  is enough to
power the observed radio nebula and its 0.3c expansion\cite{gkg+05}.

The two earlier flares would have been detectable by existing instruments only
within $\sim8$ Mpc, and, therefore, it was not previously thought that such
flares could be the source of the short, hard GRBs. The main spike of the
December 27 giant flare
would have resembled a short, hard GRB had it occurred within 40 $d_{15}$ Mpc,
encompassing even the Virgo cluster. Magnetar formation rate is expected to
follow the star formation rate, which is (for $z=0$) $1.3\pm0.2\;{\rm
M_\odot\;yr^{-1}}$  in our Galaxy\cite{kenn89} and
$0.013^{+0.02}_{-0.007}\;{\rm M_\odot\;Mpc^{-3}\;yr^{-1}}$ averaged over
intergalactic scales\cite{Gal95}. 
This suggests that the Burst And Transient
Source Experiment ({\it BATSE}) onboard the Compton Gamma Ray Observatory would
have triggered on such events as short GRBs at a rate of 
$N_{\rm BATSE} = 80(\dot{N}_{\rm gal,GF}/0.03\;{\rm yr}^{-1})d_{15}^3\;{\rm yr}^{-1}$, 
to be compared to an estimate of the $4\pi$sr {\it BATSE} rate of about $150\;{\rm
yr^{-1}}$. Here, $\dot{N}_{\rm gal,GF}$ is the average rate of giant flares in
the Galaxy similar to the Dec.~27 event.  The observed isotropic distribution
of short BATSE GRBs on the sky and the lack of excess events from the direction
of the Virgo cluster suggests that only a small fraction ($\la 0.05$) of
these events can be SGR giant flares within $\la 40\;$Mpc. This implies
either: $d \la 7$\,kpc;  $N_{\rm gal,GF}\la (3)\times 10^{-3}\;{\rm yr^{-1}}$ on
average for a Galaxy like our own; 
or the luminosity distribution includes even larger SGR flares that
can be seen at a greater distance\cite{eic02}.
One possible distinction of these from the
classic GRB population may well come from their radio observations, as their
radio afterglows should not be detectable beyond $\sim$1 Mpc.
The fraction of SGR events among what are now classified as short GRBs may not
be predominant, but it should be detectable.  This will be testable
with future \swift\ observations.


\newpage

\noindent
Correspondence should be addressed to {palmer@lanl.gov}
\medskip

\noindent
Supplementary Information accompanies the paper on www.nature.com/nature.

\smallskip

\noindent{\bf Acknowledgements}

\noindent
The authors acknowledge support from NASA (PMW, ERR, BMG);
the German-Israeli Foundation (YL);
NWO (RAMJW); DOE (JG); and the Israel-US BSF, the Israel
Science Foundation, and the Arnow Chair of Theoretical Physics
(DE).

\newpage

\begin{figure}[!p]
\centerline{\psfig{file=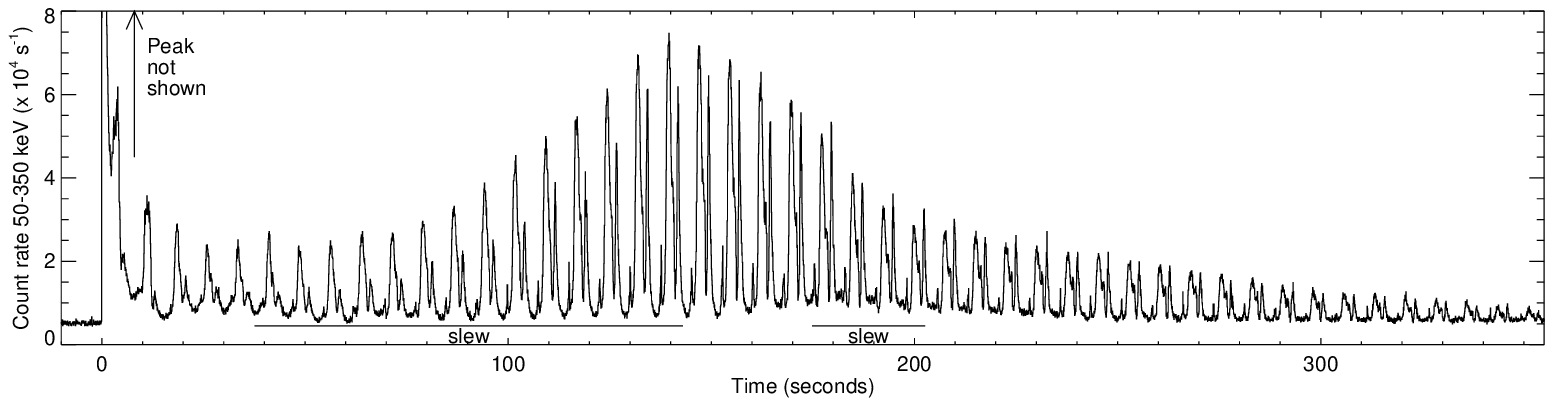,width=6.5in,angle=0}}
\bigskip
\bigskip


\noindent{{\bf Figure 1a:} The SGR spike and tail light curve from the 
Burst Alert Telescope (BAT) on \swift\ at measured energy $>$50\,keV (64 ms bins). 
Although BAT was pointed $105^{\circ}$
away from the SGR at the time of the main spike, 
it recorded $\gamma-$rays above 60\,keV passing
through and scattering within the spacecraft body and instrument shielding. 
As part of a
pre-planned observing schedule, \swift\ slewed to observe a different source
shortly after the main peak, reaching a steady pointing direction $61^{\circ}$ from
the SGR at 143 seconds. The spacecraft re-orientation improved the detection
efficiency of the SGR, visible as an apparent (not intrinsic) rise in the light
curve to a peak at 140 s.
This is followed by a second slew to $67^{\circ}$.
}

\label{}
\end{figure}

\begin{figure}[!p]
\centerline{\psfig{file=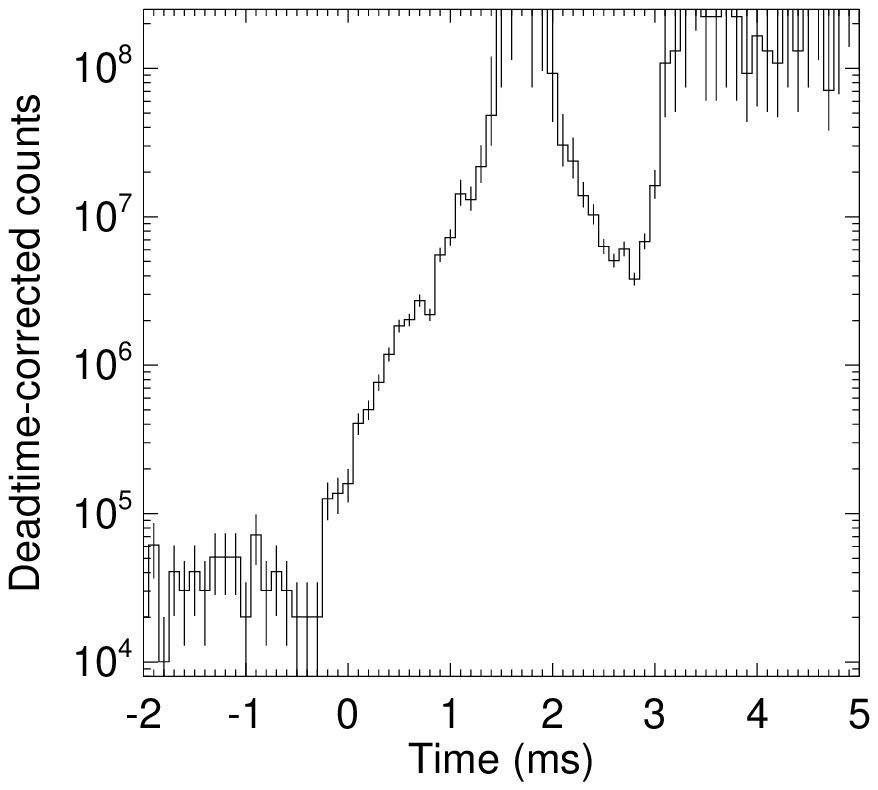,width=6.0in,angle=0}}
\bigskip
\bigskip

\noindent{{\bf Figure 1b:} BAT deadtime-corrected count rate (all energies) 
during the complex leading edge of the main spike.  (Note that horizontal
scale is $10^4$ larger than {\bf\textit{1a}}.)
Uncertainties in the deadtime correction (discussed in {\it Supplementary Methods})
make corrected count rates increasingly unreliable above $5\times10^7$ counts per second.
Time
bins of 100$\mu$s are equivalent to the light-crossing-time of a neutron star
diameter. More detailed lightcurves are shown in the {\it supplementary figures} section.}

\label{}
\end{figure}

\begin{figure}[!p]
\centerline{\psfig{file=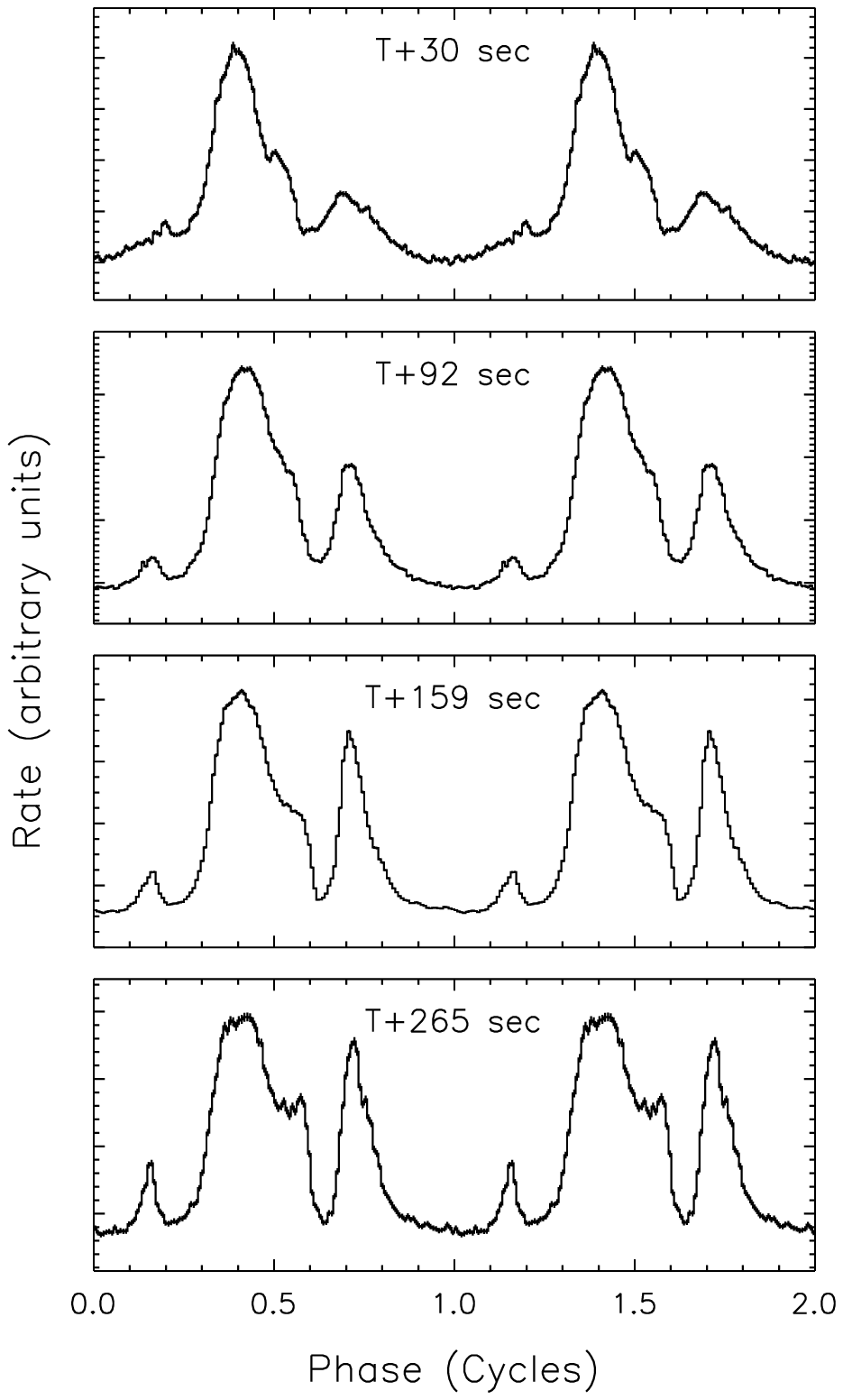,width=3.5in,angle=0}}
\bigskip

\noindent{{\bf Figure 2:} The pulse profile evolution of the magnetar \sgra\
during the giant flare of 2004 December 27.  Time through the flare increases
from top to bottom.  Each panel displays the pulse profile folded over four
pulse cycles at one of four different time intervals during the flare.  The times
denoted at the top of each panel indicate the midpoint of the interval relative
to the onset of the flare. During the first half of the tail ($\sim$170 s),
the peak centered at phase 0.7 grows in amplitude as the primary peak fades
until the two are nearly equal in height.  Thereafter, the two peaks decay in
lockstep while the relative amplitude of the third peak at phase 0.2
increases.  Overall, the pulse profile becomes less sinusoidal during the
course of the flare, that is, the power in the higher harmonics increases
relative to the power at the fundamental frequency, opposite to what was seen
in \sgrb\ during the August 27 flare\cite{fhdt01}.  The phases of the \sgra\
pulse peaks remain fixed, indicating a finalized 
magnetic geometry, and supporting the notion that after the first spike no new
magnetic energy is released, and only the trapped fireball energy leaks out. The individual pulses throughout the tail can be seen in {\it Supplementary figure} 1.}

\label{}
\end{figure}

\noindent{\LARGE \bf Supplementary Methods}

\vspace{0.15in}

\noindent{\bf SOPA and ESP peak fluence determination}

Each Synchronous Orbit Particle Analyzer (SOPA\cite{sopa}) consists of
three identical particle telescopes that function as $\Delta E$--$E$
particle spectrometers.  Energy depositions in a 3000 $\mu$m thick,
0.25 cm$^2$ silicon $E$-sensor that occur without a finite $\Delta E$
signal are identified as electrons and pulse-height analyzed into 10
deposited energy channels. To provide accurate measurements of large
particle fluxes, each SOPA telescope has a small field of view
(11$^{\circ}$) and fast readout electronics (5.5 $\mu$s deadtime). These
characteristics also make the SOPA electron channels ideal for intense
$\gamma$-ray fluxes that would overwhelm larger detectors designed for
maximum sensitivity to weak signals.
The 10 SOPA electron channels cover the deposited-energy range 45\,keV
to $>$1.5\,MeV. These data are continuously multiplexed over
individual telescopes with a cadence and resolution of 0.160\,s. The
15 SOPA telescopes on the five satellites point in different
directions, and the phase of the sampling time varies from satellite
to satellite, providing diversity in observing conditions that
allows an estimate of systematic errors by comparing
measurements across instruments.

The SOPA data for each well-placed detector were deadtime-corrected
(with typical deadtimes of $\sim50$\%) and fit to a spectral form
consisting of a power law with an exponential cutoff (a form commonly
used to fit gamma-ray burst and SGR spectra).  For some of the
detectors, the intense portion of the burst was split between two
160\,ms time bins, in others the burst was almost entirely contained
in a single time bin. This implies that the brightest part of the main spike
was less than, but comparable to, 160\,ms.

The derived spectra from the spike have `temperature' parameters
(determining the exponential cutoff energy) of $kT \sim 500$\,keV with
a spread among instruments of $\sim$100\,keV, 
and power-law index parameters of $-0.2$ with a
spread of 0.3.   
The exponential cutoff is well below SOPA's
upper energy limit.
An example of such a fit is shown
in Supplementary Figure 3.
The 1-timebin or 2-timebin integrated fluxes for each
SOPA instrument gives $5 \ergcms \times 0.160 s$, or a fluence of $0.8
\ergcm$ above 45\,keV. The spread on the derived fluence parameter
across SOPAs is $0.05 \ergcm$, which we take as a measure of the
systematic uncertainty. (The statistical uncertainty is miniscule
compared to the systematics.)

This flux is much higher than our experience with the previous two
giant SGR flares.  For perspective, the light from the full moon has
an energy flux of $3 \ergcms$.  Therefore, verification from
other instruments is desired to add to the measurement's
credibility.

The Energy Spectrometer for Particles (ESP\cite{esp}) is a
higher-energy-range instrument on the same five satellites as SOPA (a
pair of ESPs flies on each satellite).  Each is a
scintillator-photomultiplier combination, with the scintillator being a
$3.81\ {\rm cm\ dia.} \times 3.81\ {\rm cm}$ bismuth germanate cylinder
completely surrounded by 0.63-cm thick plastic scintillator as a
phoswich. The operation of this instrument includes the measurement of
the anode current, sampled every 128 ms 
(but with an effective integration time of a few ms). 
These anode currents have
been calibrated on the ground by exposure to a $^{60}$Co $\gamma$-ray
source at flux levels up to 360\ergcms.  The highest anode currents
currents measured during the superflare correspond to a peak flux
of 6 \ergcms, and so are consistent with the SOPA
measurements.

Preliminary results reported\cite{rhessi} for the RHESSI telescope
give lower limits on the fluence $>0.3$\ergcms\ over a
3\,keV-15\,MeV energy range, and $>0.1$\ergcms above 65\,keV.
These lower limits are consistent with our (somewhat higher)
measurements.

\noindent{\bf BAT tail fluence determination}

The Burst Alert Telescope (BAT)\cite{bat} on \swift\ is a wide (1.5
sr) field of view (FOV) coded aperture telescope designed to rapidly
identify and locate Gamma Ray Bursts (GRBs). The detector array is
composed of 32768 individual $4 \times 4 \times 2\ {\rm mm}^3$ CdZnTe
crystals, each with its own ampifier-discriminator chain, yielding a
5200 cm$^2$ position-sensitive detector plane with 4 mm resolution and
a nominal 15-350 keV energy range.

As \swift\ was oriented at the time of the main spike, the SGR
illuminated the detector at an angle of 105$^{\circ}$ from the center of the
field of view.  At this angle, the $\gamma$-rays passed through or were
scattered by the body of the spacecraft, the tube of either the X-Ray
Telescope (XRT) or UV and Optical Telescope UVOT (depending on
detector location) and a graded-Z shield behind the detector plane.
After the first and second slew, giving angles of 61$^{\circ}$ and 67$^{\circ}$,
respectively, the $\gamma$-rays no longer had to pass through the
spacecraft body (although they still encountered the XRT or UVOT and
graded-Z shielding on the sides of the BAT before reaching the
detector).

Using an instrumental and spacecraft mass model and Monte Carlo
methods, we generated Detector Response Matrices (DRMs) for the
105$^{\circ}$ and 67$^{\circ}$ orientations, allowing us to use the XSPEC
package to derive fluxes and spectra from count rates.  The effective
area of the instrument in the 105$^{\circ}$ orientation is a few hundred
cm$^2$ at photon energies above 100 keV, dropping to $\sim$10\,cm$^2$
around 60 keV and insignificant below 45 keV.  In the 67$^{\circ}$
orientation the effective area was roughly twice that for 105$^{\circ}$ at
photon energies of 60 keV and above, and maintained a small but
appreciable effective area of 3~cm$^2$ at our 15 keV threshold.  (In
that orientation, low energy photons could pass through
the coded aperture mask and scatter off the innermost (copper) layer
of the shield directly into the detector.)  Strong instrumental
fluorescence lines are predicted by the Monte Carlo, and are indeed
the most obvious features of the measured spectrum throughout the tail
emission.

One of the data products that BAT generates in response to a GRB or
other transient is a listing of every detected photon, including the
measured energy deposition and the time to $100 \mu$s resolution.  For
transients that are not located ({\it e.g.\/}, as in the case of the
SGR, when they are not in the BAT's imaged field-of-view), 20 seconds
of such data is returned.  This is sufficient to produce the high time
resolution plot of Figure 1b, but did not include most of
the tail emission.  However, continuous data is available as count
rates in 4 energy bins at 64 ms resolution.

We integrated the count rates in the two highest energy bins (50--100
and 100--350\,keV deposited energy) to get the background-subtracted
two-channel `spectra' in three time intervals: before, during, and
after the first slew.

These spectra were then fit with XSPEC using two spectral forms:
an unbroken power-law (PL) and an Optically Thin Thermal Bremsstrahlung
(OTTB).  The first interval was fit using the 105$^{\circ}$ DRM, the third interval was
fit using the 67$^{\circ}$ DRM, and the second interval, the slew, was fit using
both.  As each individual fit is constrained (two free parameters fitting two spectral bins),
and because systematic uncertainties are large while uncertainties in
counting statistics are practically non-existent,
$\chi^2$ fit quality is inapplicable and error estimates must come from
arguments such as consistency levels among fits.

The PL fits gave large spectral indices ($-5$ -- $-7$), which is usually an indicator that
an exponential cut-off is required by the fit.  The OTTB (a spectral form that includes
an exponential cut-off) gave fits with temperature parameters in the 15-35\,keV range,
consistent with previous measurements of giant flare tails.  The fit temperature
is unreliable, based solely on the `color' (two-bin count ratio) at energies
well above the peak, but the consistency gives qualitative support to 
the use of this spectral form.

The calculated energy fluxes based on these two spectral forms 
were consistent within 20\% over 60--350\,keV.  Fits
made over the slew interval with the two DRMs varied in calculated
fluence by a factor of 2, consistent with the change in effective area
over this time interval.  Integrating the fluences over all three intervals
(using the average fluxes of the two DRMs during the slew interval)
gave $0.93\times10^{-3}$ and $1.1\times10^{-3}$ \ergcm\ for the OTTB and PL,
respectively.

Based on the obvious systematic uncertainties in this procedure,
we adopted a systematic error estimate of 50\%.
This is much larger than the discrepancy between the two
models' fits, but many of the systematics of fitting
two-parameter models with two measurements are correlated.

In addition, the width of the pulses in each cycle are narrow, indicating
that the tail emission is substantially beamed.  Measurements at Earth
show 3 or 4 peaks, but observers at other latitudes relative to the
neutron star would see a different pulse pattern, with a different overall
intensity.  ({\it E.g.}, the August 27 1998 tail from \sgrb\ showed 4 main peaks
with 8 distinct maxima per cycle.)  
Thus, even perfect measurements would provide
limited advantage in determining the total energy radiated during the tail.

Averaging the fluences derived
from the two spectral forms yielded our stated fluence of $1.0(5)\times10^{-3}$\ergcm.

\vspace{0.15in}

\noindent{\bf BAT deadtime correction}

BAT's 32,768 individual CdZnTe detectors are grouped into 256 units,
called 'sandwiches'.  Each sandwich contains 128 detectors feeding an integrated
circuit (the XA-1) that contains 128 amplifier-discriminator chains and a demultiplexer.  
When a detector
absorbs a $\gamma$-ray, the signal is amplified, then detected by the discriminator,
which triggers circuitry to demultiplex that signal to the XA-1 analog output (giving 
a measure of the deposited energy) and to indicate which detector was struck.
While this data is digitized by circuitry associated with the sandwich, further
signals from the 128 detectors of that sandwich are ignored.  This `deadtime'
has been measured at $t_{dead} = 98(3) \mu s$.
Thus the maximum combined rate at which the 256 sandwiches of 
BAT can detect photons is the saturation rate
$R_{sat} = 256 / t_{dead} = 2.61(8) \times 10^{6} s^{-1}$.

If the BAT is measuring photons at a rate of $r_{meas}$, then
the average sandwich is dead for $r_{meas}/R_{sat}$ of the time,
so the detector rate can be deadtime corrected by dividing by the
livetime fraction:
$r_{corr} = r_{meas} / (1 - r_{meas}/R_{sat})$.
As $r_{meas}$ approaches $R_{sat}$ the uncertainty in $t_{dead}$
makes $r_{corr}$ poorly determined.  Thus the deadtime correction from BAT
measured count rate to photon rate is currently highly uncertain above $r_{corr}\ga20 R_{sat}$.
Above $r_{corr}\ga30 R_{sat}$ the formal error range includes
infinite flux as the deadtime is consistent with 100\%.
The peak fluxes from SOPA indicate that the true photon rate on the BAT
detector was $\sim10^3 R_{sat}$, and so
even a much more accurate $t_{dead}$ determination would have resulted
in a a poor estimate of  $r_{corr}$ at the peak.  
Thus SOPA and ESP are needed to provide the peak measurements
that BAT cannot.

\newpage

\begin{figure}[!p]
\centerline{\psfig{file=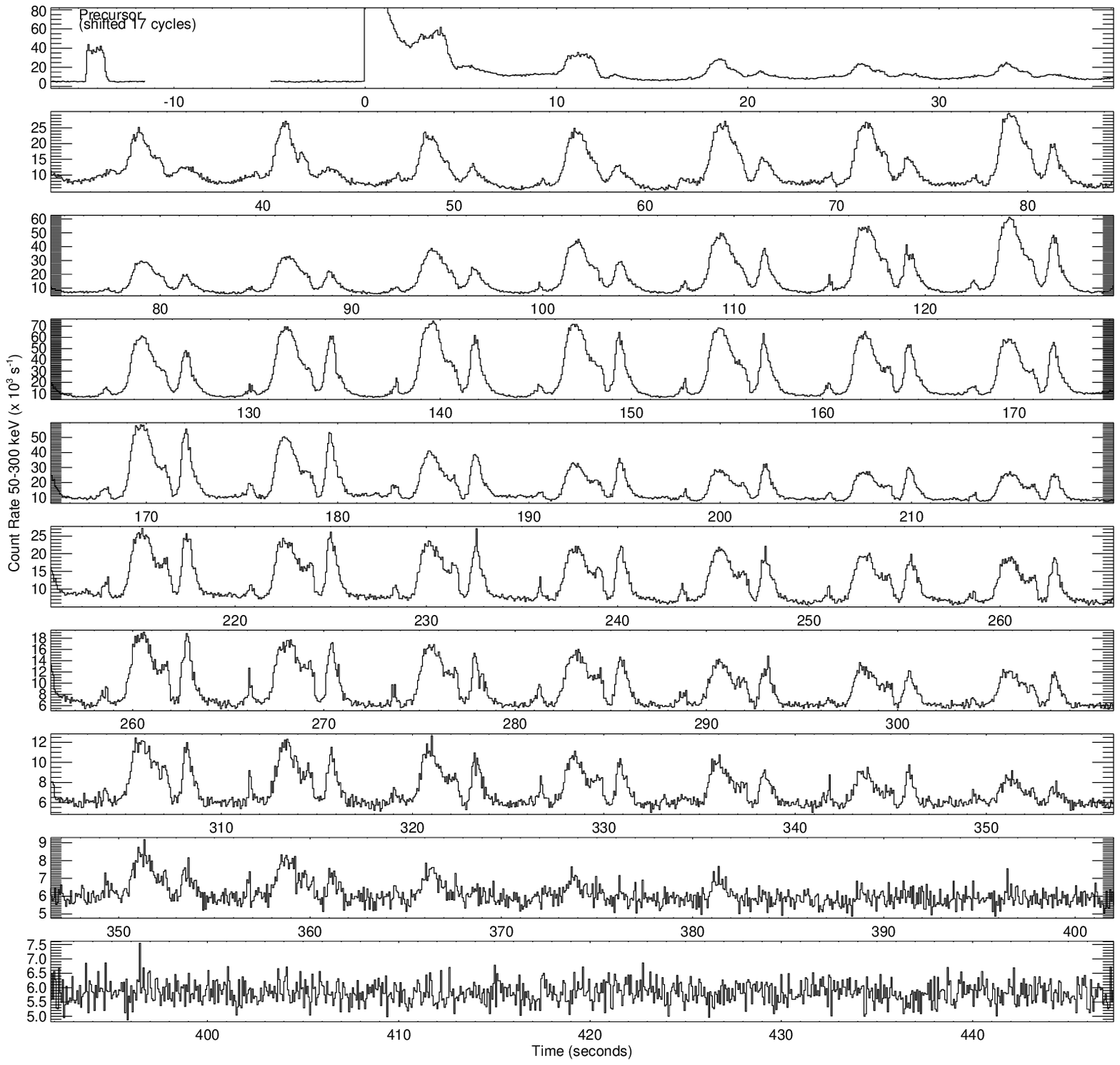,width=6.5in,angle=0}}
\bigskip
\bigskip

 {{\bf Supplementary Figure 1:}
BAT lightcurve of the precursor and tail of the giant flare with 64 ms resolution. 
Successive plots are phase aligned at the 7.56 s neutron star spin frequency,
with one cycle of overlap from plot to plot.  The precursor, which occurred 
142\,s before the spike, is plotted at
the appropriate spin phase.  
 }
\end{figure}

\begin{figure}[!p]
\bigskip
\bigskip

\centerline{\psfig{file=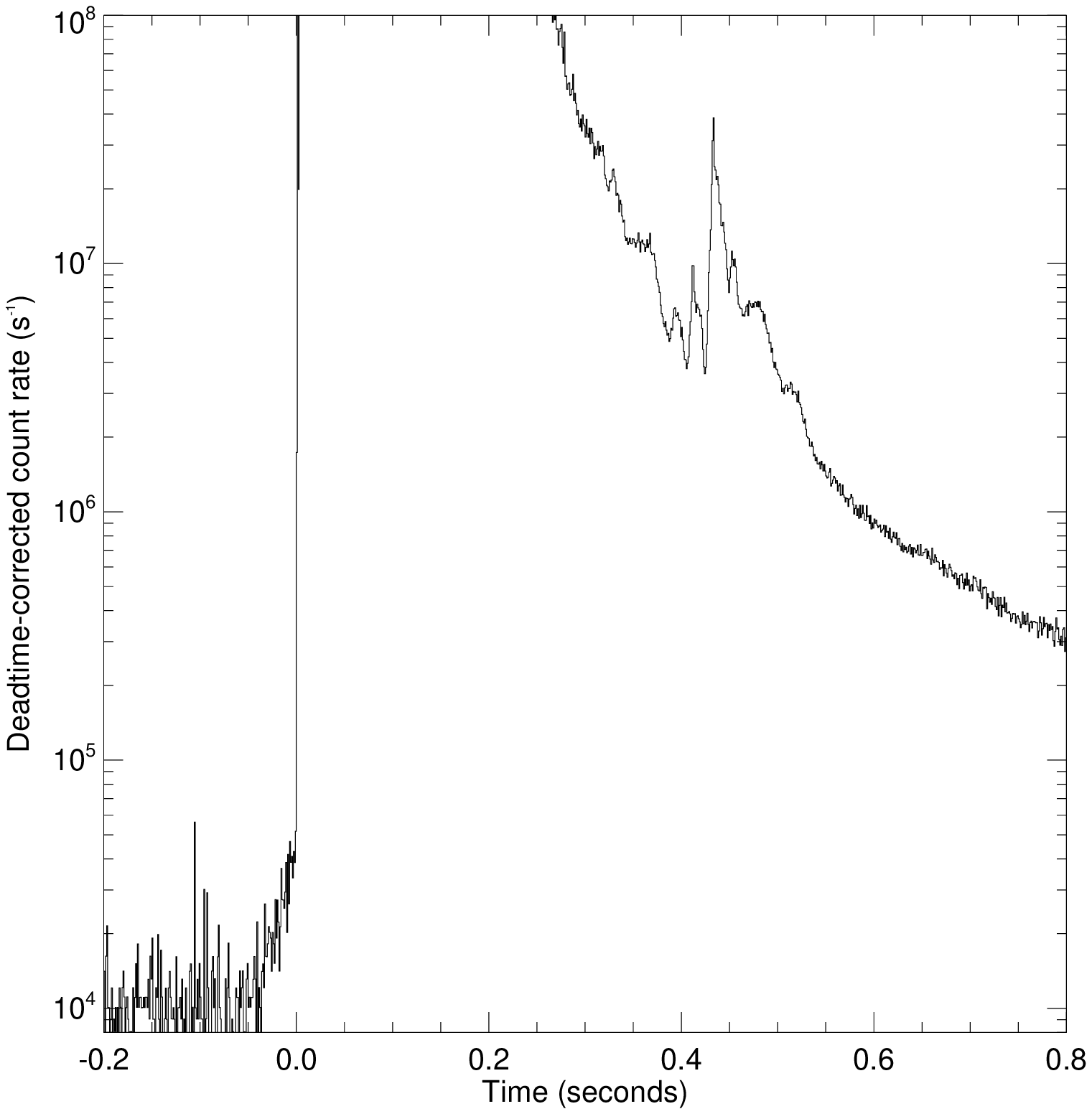,width=6.5in,angle=0}}
 {{\bf Supplementary Figure 2:} 
BAT deadtime-corrected lightcurve leading up to and during the decline of the main spike,
with 1 ms resolution.  
Corrected rates above $5 \times 10^7$ have large uncertainties in the
deadtime correction.
  }
\end{figure}

\begin{figure}[!p]
\bigskip
\bigskip
 
 \centerline{\psfig{file=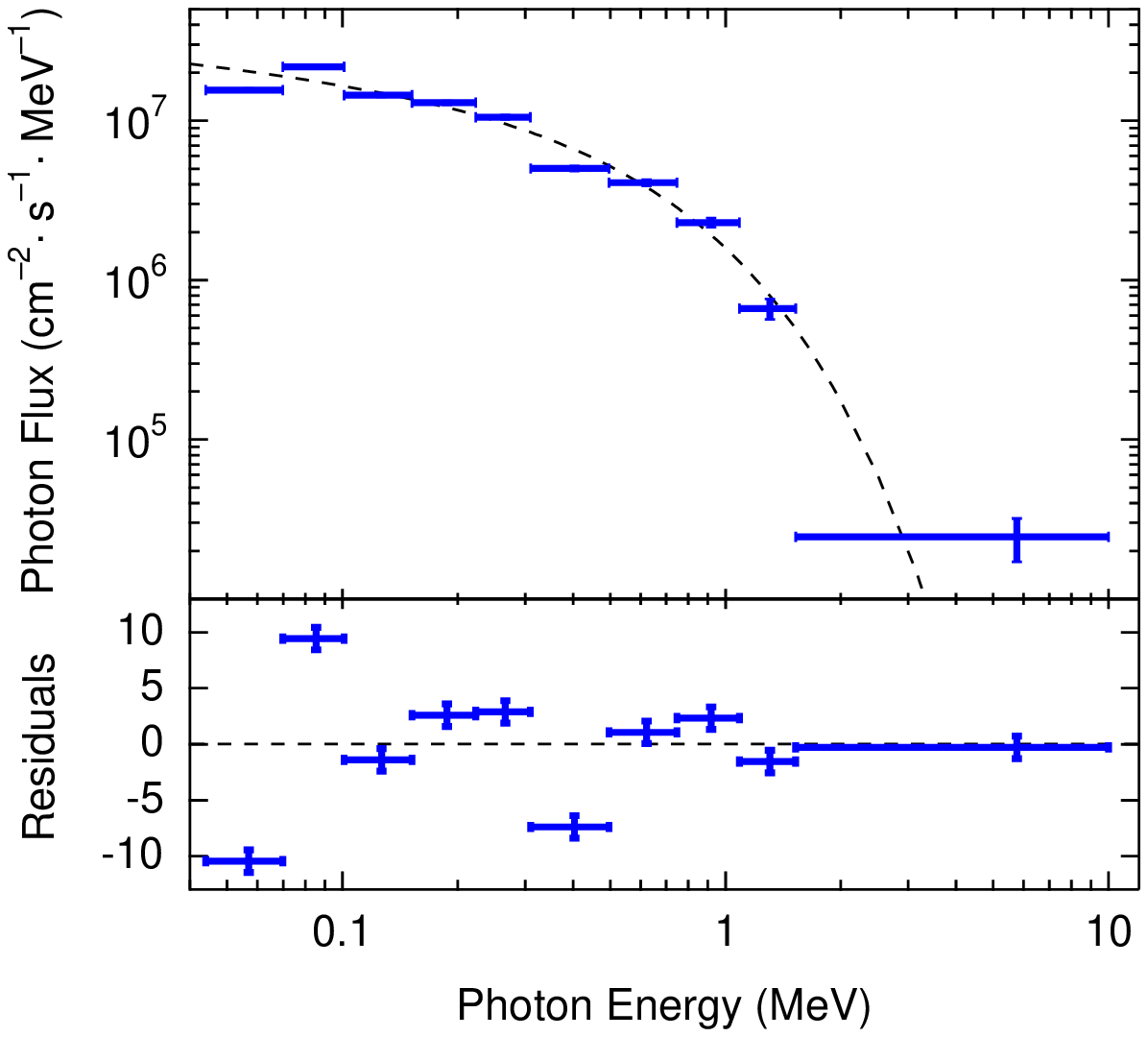,width=6.5in,angle=0}}
 {{\bf Supplementary Figure 3:} A spectral fit to SOPA data.
This is an example of a spectral fit to data from one of the SOPA detectors
to the peak 0.160 s of the main spike.
The dashed line is the spectral model in photons, with a low energy power law 
 $\alpha$= -0.2 and an exponential cut-off kT = 0.48 MeV. }
\end{figure}

\end{document}